\documentclass[letterpaper,american,aps,prl,superscriptaddress,showpacs,floatfix,twocolumn]{revtex4}
\pdfoutput=1
\usepackage{mathptmx}
\usepackage[T1]{fontenc}
\usepackage[latin2]{inputenc}
\usepackage{graphicx}
\usepackage{babel}

\begin{document}

\title{Dirac cones and minigaps for graphene on Ir(111)}

\author{I. Pletikosić}
\author{M. Kralj}
\email{mkralj@ifs.hr}
\author{P. Pervan}
\affiliation{Institut za fiziku, Bijenička 46, 10000 Zagreb, Croatia}

\author{R. Brako}
\affiliation{Institut Ruđer Bošković, Bijenička 54, 10000 Zagreb, Croatia}

\author{J. Coraux}
\author{A. T. N'Diaye}
\author{C. Busse}
\author{T. Michely}

\affiliation{II. Physikalisches Institut, Universität zu Köln, Zülpicher Straße 77, 50937 Köln, Germany}
\begin{abstract}
Epitaxial graphene on Ir(111) prepared in excellent structural quality
is investigated by angle-resolved photoelectron spectroscopy. It clearly
displays a Dirac cone with the Dirac point shifted only slightly above
the Fermi level. The moiré resulting from the overlaid graphene and
Ir(111) surface lattices imposes a superperiodic potential giving
rise to Dirac cone replicas and the opening of minigaps in the band
structure. 
\end{abstract}


\maketitle
Graphene has recently become a material of increasing scientific interest
\citep{Geim:NM2007,CastroNeto:RMP2008}. Its honeycomb structure composed
of two equivalent triangular carbon sublattices has important consequences
for the dynamics of the charge carriers. The $\pi$ and $\pi^{*}$
bands of free-standing graphene are conical in the proximity of the
Fermi energy, with the vertices touching exactly at the Fermi level.
This allows a mapping of their behavior to a model of massless fermions
obeying the Dirac equation, leading to a plethora of new, observed
or predicted phenomena, such as anomalous half-integer quantum Hall
effect or the Klein paradox (see e.g. ref. \citep{Geim:NM2007}).
The Dirac character and the large mobility of charge carriers in graphene
\citep{Bolotin:PRL2008}, along with the ability to manipulate conduction
through field effect \citep{Novoselov:S2004} or doping \citep{Bostwick:NP2007}
makes graphene potentially a material for future electronics.

Most of the experimental work on the electronic structure of supported
graphene has been performed with graphene on SiC \citep{Berger:S2006,Ohta:S2006,Bostwick:NJP2007,Rotenberg:NM2008,Zhou:NM2008,Bostwick:NP2007}.
Early research related to graphene on transition metal surfaces \citep{Hagstrom:PRL1965,May:SS1969,Oshima:JP1997,Gall:IJMPB1997}
has received renewed interest in the last years as epitaxial graphene
layers of high structural quality can be grown \citep{Dedkov:PRL2008,Grueneis:PRB2008,Marchini:PRB2007,deParga:PRL2008,Sutter:NM2008,Coraux:NL2008,NDiaye:NJP2008}.
Moreover, the electronic interaction of graphene with a metal is not
only of fundamental interest. Recent calculations appear to imply
that graphene which interacts weakly with a metal will result in a
favorable contact with high transmission \citep{Nemec:PRB2008}. The
charge carrier properties of graphene on a metal have been studied
on only two substrates, Ni(111) \citep{Dedkov:PRL2008,Grueneis:PRB2008}
and Ru(0001) \citep{Marchini:PRB2007,deParga:PRL2008,Sutter:NM2008}.
The $\pi$-bands of graphene strongly hybridize with Ni(111) \citep{Grueneis:PRB2008},
while the small bonding distance of the first layer of graphene and
the absence of vibrational/electronic signatures characteristic of
graphene indicate the same on Ru(0001) \citep{deParga:PRL2008,Sutter:NM2008}. 

For the epitaxially grown graphene where the substrate has a different
lattice constant, the stiffness of graphene leads to the formation
of structures with large superperiodicity. Simulations within the
effective-Hamiltonian formalism for graphene subjected to superperiodic
potentials suggest that graphene superlattices could be used for tuning
the propagation velocity and the density of charge carriers in graphene,
which is of practical interest in building graphene based electronic
components \citep{Park:NP2008}. The main effect of a superperiodic
potential on the band structure is the formation of minigaps at the
crossing points of a band and a backfolded band, as found, e.g., for
$\mathrm{Si(111)\textrm{-}\sqrt{21}\times\sqrt{21}\textrm{-}(Ag+Au)}$
\citep{Crain:PRB2002}. However, minigaps were neither observed in
graphene on SiC \citep{Bostwick:NJP2007} nor on Ru(0001), where they
are expected to be present \citep{deParga:PRL2008}.

Here we report on the investigation of the electronic structure of
graphene on Ir(111) by angle-resolved photoelectron spectroscopy (ARPES).
We show that graphene exhibits a Dirac cone comparable to that of
pristine graphene. Due to the moiré superstructure resulting from
the lattice mismatch between graphene and Ir(111) a corresponding
superperiodic potential exists which gives rise to the opening of
moiré-induced minigaps in the band structure.

The experiments have been performed in two ultra-high vacuum systems
for ARPES in Zagreb and scanning tunneling microscopy (STM) in Cologne
with base pressures in the low $10^{-8}\;\mathrm{Pa}$ range using
identical procedures for substrate cleaning and graphene fabrication.
Graphene was prepared on the clean Ir(111) surface by 7--15 cycles
of room temperature ethene adsorption to saturation and subsequent
thermal decomposition at 1450~K. ARPES spectra have been taken at
60~K by a Scienta SES-100 hemispherical electron analyzer with an
energy resolution of \textasciitilde{}25~meV using 21.22~eV photons
from a helium discharge source with a beam spot diameter of about
2 mm. For the parallel momentum scan the polar angle $\vartheta$
ranged from $40^{\circ}$ to $70^{\circ}$. The azimuthal angle, $\varphi$,
was changed by a wobble stick and checked by LEED spot orientation
with the precision of $\pm0.5^{\circ}$. Below we show that for a
range of azimuths within a few degrees around the $\mathrm{\Gamma\textrm{-}K\textrm{-}M}$
direction, the absolute value of $\varphi$ can be determined to a
precision better than $0.1^{\circ}$ when the measured dispersion
is fitted by the tight-binding approximation (TBA) bands of graphene
\citep{Reich:PRB2002}.

\begin{figure}
\includegraphics[width=86mm]{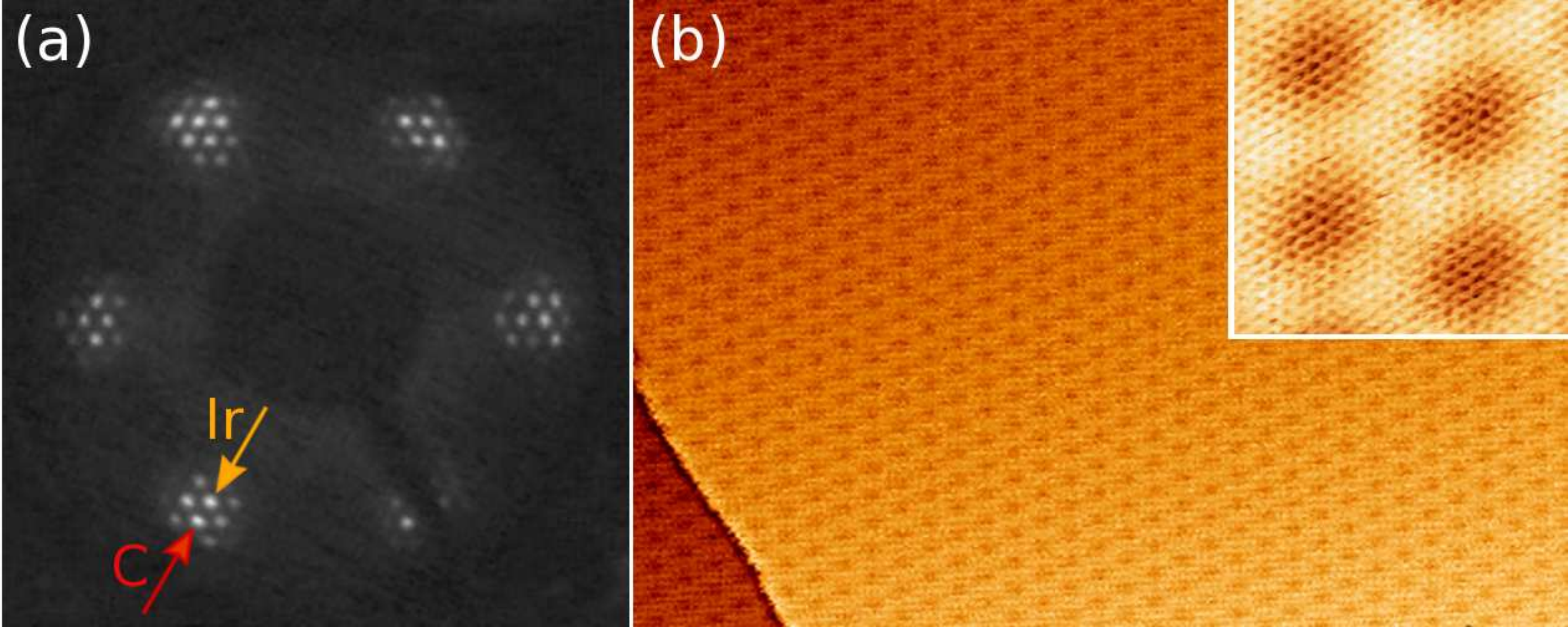} 

\caption{\label{fig:LEED-STM}(color online). (a) LEED pattern (69~eV) of
graphene on Ir(111). First order Ir spots, C spots, as well as surrounding
moiré spots are visible. (b) STM topograph ($75\:\mathrm{nm}\times50\:\mathrm{nm}$)
exhibiting moiré contrast (enhanced by an unsharp masking filter).
Tunneling conditions 0.05~V, 32~nA. Inset: $5\:\mathrm{nm}\times5\:\mathrm{nm}$
atomic resolution image of graphene lattice. Tunneling conditions
0.37~V, 32~nA.}

\end{figure}

Graphene on Ir(111) grows with graphene $[1\,1\,\bar{2}\,0]$ and
Ir $[1\,0\,\bar{1}]$ directions almost perfectly aligned. Due to
the difference in lattice constants, a (9.32\,$\times$\,9.32) moiré
with a repeat distance of 2.53~nm is formed \citep{NDiaye:NJP2008}.
The low energy electron diffraction (LEED) patterns -- indistinguishable
in appearance in our two setups -- display aligned first order diffraction
spots of the Ir surface and graphene lattices {[}arrows in Fig.\ \ref{fig:LEED-STM}(a){]}
surrounded by the spots of the moiré. The relative position of the
carbon atoms with respect to the iridium surface atoms changes within
the moiré cell, giving rise to a contrast in STM with the moiré periodicity
{[}Fig.\ \ref{fig:LEED-STM}(b){]}. The amplitude and sign of the
moiré contrast depend on tunneling voltage \citep{NDiaye:NJP2008},
indicating its largely electronic origin. From images of higher magnification
{[}inset of Fig.\ \ref{fig:LEED-STM}(b){]}, which display the moiré
contrast together with the graphene lattice, the misalignment of graphene
with respect to the substrate is accurately determined. Statistics
for \textasciitilde{}25\,\% graphene covered surface yields a $0.25^{\circ}$
scatter in orientation between the two lattices. STM imaging of the
surface covered up to 70\,\% by graphene shows flakes with lateral
dimensions ranging from several hundred nanometers to a few micrometers.
The structural perfection and insignificant orientation scatter (comparable
with the $0.1^{\circ}$ angular resolution of our electron analyzer)
of graphene are a prerequisite for the successful application of ARPES.

\begin{figure}
\includegraphics[width=86mm]{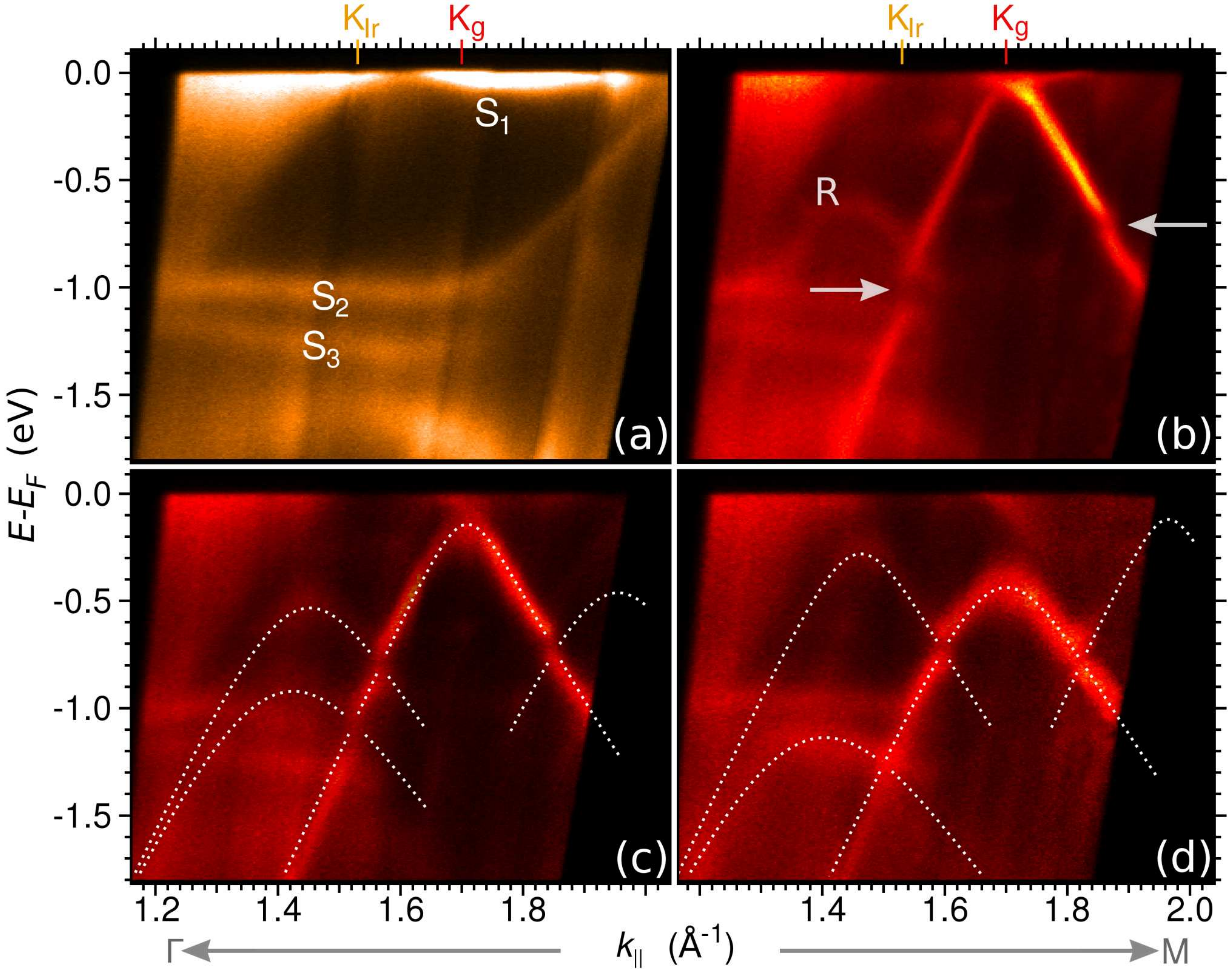} 

\caption{\label{fig:ARPES} (color online). (a) ARPES spectrum of clean Ir(111),
$\varphi=(0.5\pm0.1)^{\circ}$. The positions of K points of iridium
and graphene are marked as $\mathrm{K_{Ir}}$ and $\mathrm{K_{g}}$,
respectively. $S_{1}$--$S_{3}$ are surface states. (b) ARPES spectrum
of Ir(111) covered by graphene along the same azimuth as in (a). Horizontal
arrows denote the minigap in the primary Dirac cone. A visible replica
band is labeled as R. Detailed analysis in \citep{EPAPS}. (c),(d)
ARPES spectra for $\varphi=(1.4\pm0.1)^{\circ}$ and $\varphi=(3.0\pm0.1)^{\circ}$,
respectively. The dashed lines are TBA calculated bands for Dirac
cones located at positions 0, 1, 2, and 5 defined in Fig.\ \ref{fig:mBZ}(b). }

\end{figure}

\begin{figure*}
\includegraphics[width=170mm]{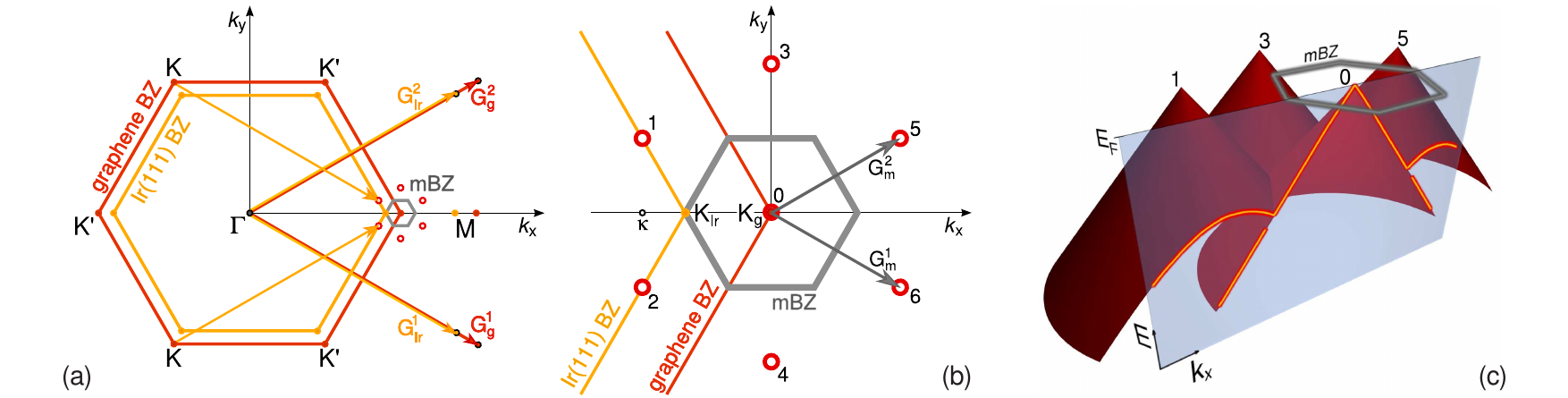}

\caption{\label{fig:mBZ} (color online). (a) Brillouin zones (BZ) of the Ir(111)
surface and graphene lattices drawn to scale. Mini Brillouin zone
(mBZ) is indicated around the K point of graphene. Open dots denote
the vertices of Dirac cones replicated by moiré reciprocal vectors,
$\mathbf{G}_{\mathrm{m}}$. (b) Blow-up of the mBZ. Central dot denotes
the center of the primary Dirac cone, while 1--6 are replicas.  (c)
ARPES experimental plane (shaded) along the $\mathrm{\Gamma\textrm{-}K\textrm{-}M}$
high symmetry direction cuts the primary and replica cones (only partially
shown). Bright lines at the intersection mark the bands observed in
Fig.\ \ref{fig:ARPES}(b).}

\end{figure*}

Due to the alignment of the graphene lattice and the Ir(111) surface
lattice their Brillouin zones (BZ) are aligned as well (cf. Fig.\ \ref{fig:mBZ}).
The spectrum of clean Ir(111) along $\mathrm{\Gamma\textrm{-}K\textrm{-}M}$
shown in Fig.\ \ref{fig:ARPES}(a) encompasses a region where the
graphene Dirac cone is expected. Conveniently, this region coincides
with an energy gap in the Ir(111) electronic structure. Three weakly
dispersing states, which we identify as surface states, are clearly
visible in the gap: $S_{1}$ near the Fermi level, and $S_{2}$ and
$S_{3}$ passing near the low edge of the gap. The ARPES spectrum
of graphene on Ir(111), shown in Fig.\ \ref{fig:ARPES}(b), contains
all the features required to discuss the essential properties of graphene:
(i) presence of a Dirac cone, (ii) marginal doping, (iii) additional
bands, and (iv) opening of minigaps. The $\pi$-band with linear dispersion
up to the Fermi level, the Dirac cone, is visible at the graphene
K-point \citep{hybrid}. The $\pi$-band spectrum is sharp, with a
full width at half maximum not larger than $0.15\;\mathrm{eV}$ and
$0.035\;\textrm{\AA}^{-1}$ in the energy and momentum distribution
curves, respectively. The Dirac cone branches have noticeably anisotropic
intensity. The measured ARPES $\pi$-band dispersion is accurately
reproduced for all measured azimuths by the tight-binding approximation
bands for the unperturbed graphene described in \citep{Reich:PRB2002},
shifted in energy in order to obtain the best fit of the position
of the main and replicated Dirac cones. According to this fit we estimate
the position of the Dirac point to be $(0.10\pm0.02)\;\mathrm{eV}$
above the Fermi energy, i.e. graphene on Ir(111) is just slightly
p-doped. As the Dirac point is not accessible with ARPES, we are unable
to judge whether a band gap opens at the Dirac point or not \citep{doping}.
However, if there is one, it ought to be smaller than twice the distance
from the Fermi energy to the estimated Dirac point, i.e. its width
must be smaller than $0.20\;\mathrm{eV}$. Besides the primary Dirac
cone Fig.\ \ref{fig:ARPES}(b) displays an additional band, a faint
replica labeled R. At the location where the band R intersects the
Dirac cone in the $\mathrm{\Gamma\textrm{-}K}$ branch {[}indicated
by a horizontal arrow in Fig.\ \ref{fig:ARPES}(b){]} a band gap
is visible, in the following referred to as minigap. A minigap is
also visible in the $\mathrm{K\textrm{-}M}$ branch in the Dirac cone
in Fig.\ \ref{fig:ARPES}(b).

In order to clarify the origin of the additional band R and the minigaps
in Fig.\ \ref{fig:ARPES}(b) we show the Brillouin zones of graphene
and Ir(111) in Fig.\ \ref{fig:mBZ}(a). The reciprocal lattice vectors
$\mathbf{G}_{\mathrm{Ir}}$ and $\mathbf{G}_{\mathrm{g}}$ of the
Ir surface and graphene, respectively, give rise to the moiré reciprocal
vectors $\mathbf{G}_{\mathrm{m}}=\mathbf{G}_{\mathrm{g}}-\mathbf{G}_{\mathrm{Ir}}$,
the same which create the LEED pattern in Fig.~\ref{fig:LEED-STM}(a).
From the dependence of the moiré contrast on tunneling parameters
\citep{NDiaye:NJP2008} it is evident that the electron potential
of graphene is modulated by the substrate. This superperiodic potential,
with the corresponding $\mathbf{G}_{\mathrm{m}}$ vectors, creates
replica bands centered at the points labeled 1--6 in Fig.\ \ref{fig:mBZ}(b)
and opens the gap in the Dirac cone along the mini BZ (mBZ) boundary.
For the $\mathrm{\Gamma\textrm{-}K\textrm{-}M}$ experimental scan
direction the cones 1 and 2 give rise to the hyperbolic band with
the maximum centered at $\kappa$ {[}schematically shown in Fig.\ \ref{fig:mBZ}(b)
and \ref{fig:mBZ}(c){]}, visible as the band R in Fig.\ \ref{fig:ARPES}(b).
As we turn the scan direction off the $\mathrm{\Gamma\textrm{-}K\textrm{-}M}$,
we move away from the vertex of the primary Dirac cone. Its ARPES
cut gets lower in energy, while the opposite happens for the replicas
whose vertex is approached, as exemplified for two azimuths in Figs.\ \ref{fig:ARPES}(c)
and (d).

\begin{figure}
\includegraphics[width=66mm]{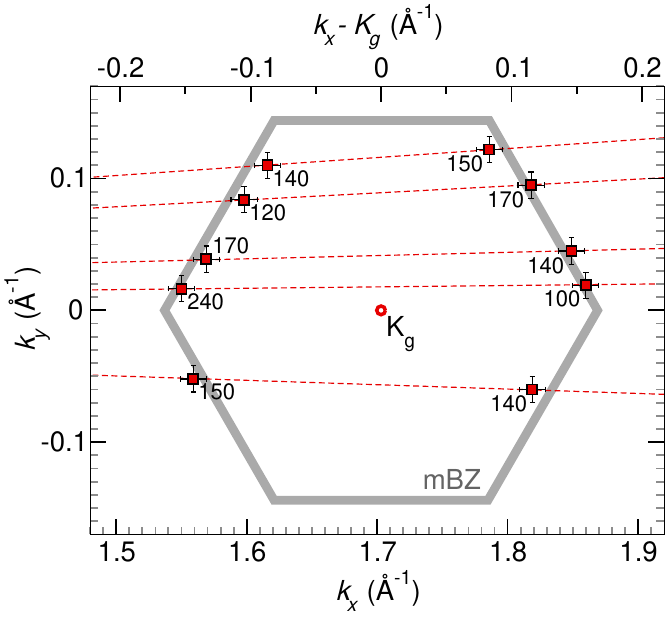}

\caption{\label{fig:gaps} The minigaps determined from ARPES scans along the
directions indicated by dashed lines fall on the edge of the mini
Brillouin zone. Their widths are annotated in meV. The dashed lines
correspond to $\varphi=-1.9^{\circ}$, $\varphi=0.5^{\circ}$, $\varphi=1.4^{\circ}$,
$\varphi=3.0^{\circ}$, and $\varphi=3.9^{\circ}$.}

\end{figure}

Replica cones in epitaxial graphene may arise as a final state effect
through surface umklapp, involving the reciprocal Ir surface lattice
vectors \textbf{$\mathbf{G}_{\mathrm{Ir}}^{1}$} and \textbf{$\mathbf{G}_{\mathrm{Ir}}^{2}$
}{[}cf. Fig.\ \ref{fig:mBZ} (a){]} in a momentum change of the photoelectron
diffracted at the surface. This mechanism was put forward in explaining
the nature of replicas in graphene on SiC(0001) where, contrary to
our observation, the replica cone intensity disappeared at room temperature
\citep{Bostwick:NJP2007}. Consistent with this interpretation no
gaps at the intersection of primary and replica cones were observed
\citep{Bostwick:NJP2007,Ohta:S2006}. Thus, the presence of superperiodic
contrast in STM topographs of graphene on SiC(0001) is then to be
interpreted as a purely geometric height modulation \citep{Mallet:PRB2007}.
Though we cannot exclude final state contributions to the intensity
of the replica cones, the surface umklapp alone cannot create the
minigap reported here. Therefore, we consider the presence of the
minigap as an evidence for an initial state effect in ARPES, i.e.
the existence of a moiré periodic potential. The analysis of ARPES
spectra reveals that the minigap is indeed located on the Bragg planes
bounding the mBZ (Fig.\ \ref{fig:gaps}). The measured gap width
is in the range 0.1--0.2 eV. Consequently, the amplitude of the moiré
potential is of the order of 0.05--0.10~eV, small compared to the
$\pi$-band width. No closing of the minigap was observed. Note that
a model using a long-wavelength superperiodic potential which does
not break the symmetry of the two graphene sublattices opens a minigap
which vanishes at specific points of the boundary of the superstructure
BZ, but does not open the gap at the Dirac point \citep{Park:NP2008}. 

The symmetry breaking as a consequence of the inequivalence of two
carbon atoms within each unit cell of graphene ought to imply opening
of a band gap at the Dirac point \citep{Zhou:NM2007}, and lifting
of the complete asymmetry of the ARPES intensity \citep{Bostwick:NJP2007,Daimon:JES1995}.
A bandgap due to symmetry breaking has been clearly observed in the
bilayer and few layer graphene on SiC(0001) \citep{Bostwick:NJP2007,Ohta:S2006}
but it remains under debate for monolayer graphene on the same substrate
\citep{Rotenberg:NM2008,Zhou:NM2008}. A detailed analysis of the
moiré of graphene on Ir(111) shows that the atoms in the two sublattices
of graphene are locally and globally inequivalent \citep{NDiaye:NJP2008}.
The global inequivalence reveals itself through the fact that Ir clusters
formed on graphene \citep{NDiaye:PRL2006,Feibelman:PRB2008} bind
only to the areas of the moiré where the carbon atoms of one sublattice
sit atop of substrate iridium atoms. Our ARPES spectra show visible
asymmetry of the photoemission intensity, which along with the estimate
of a small gap at the Dirac point ($<0.2\,\mathrm{eV}$) indicates
that the effects of the symmetry breaking for graphene on iridium
are small. 

As one of the most prominent features, the Dirac cone of graphene
on Ir(111) shows no sign of hybridization with substrate electronic
bands. Consistently, density functional theory (DFT) calculations
taking into account the large supercell of graphene on Ir(111) suggest
a weak bonding of graphene \citep{Feibelman:PRB2008,NDiaye:PRL2006}.
In particular, DFT calculations yield a large average graphene --
Ir(111) separation of $0.34\;\mathrm{nm}$ \citep{Feibelman:PRB2008},
which also explains the defect free growth of the graphene flakes
over step edges \citep{Coraux:NL2008}. Our measurements suggest that
the weakness of bonding can be ascribed to the dearth of metallic
states of appropriate symmetry (i.e., the existence of the band gap)
around the K-point of the surface BZ of Ir(111). Marginal p-doping
comes out as a fortuitous match of several system parameters: proper
graphene -- Ir(111) distance, work function difference, and weak chemical
bonding \citep{Giovannetti:PRL2008}.

Our experiments single out graphene on Ir(111) as a system with unique
properties as compared with other epitaxially grown graphene systems.
In distinction to graphene on Ni(111) \citep{Grueneis:PRB2008,Dedkov:PRL2008}
and Ru(0001) \citep{Marchini:PRB2007,deParga:PRL2008,Sutter:NM2008}
it is a weakly interacting graphene. In distinction to graphene on
SiC(0001) \citep{Berger:S2006,Bostwick:NJP2007,Bostwick:NP2007,Mallet:PRB2007,Ohta:S2006,Rotenberg:NM2008,Zhou:NM2007,Zhou:NM2008}
and two-layer graphene on Ru(0001) \citep{Sutter:NM2008} it is a
simple system with no carbon rich intermediate layer necessary to
passivate the substrate. It is this simplicity, together with its
unparalleled structural quality, which makes it an attractive candidate
for model experiments. To give an example, for graphene on Ir(111)
there is a straightforward method to vary the strength of the moiré
periodic potential through ordered adsorption of small metal clusters
of well defined size into the moiré \citep{NDiaye:PRL2006}, allowing
one experiments on charge carrier manipulation through periodic potentials
\citep{Park:NP2008}.

In conclusion, our ARPES measurements show that the electronic bands
of graphene on Ir(111) near the Fermi level have the form of the Dirac
cone of pristine graphene, only slightly shifted to lower binding
energies due to a marginal p-doping by the substrate. The periodic
potential associated with the moiré gives rise to the formation of
replica cones and minigaps at the Bragg planes between the primary
and replica cones. The simplicity of the system and its desirable
structural and electronic properties make it interesting for fundamental
research and a candidate for graphene-metal contacts in potential
applications of graphene in electronics. 
\begin{acknowledgments}
M. K. and P. P. acknowledge fruitful discussions with T. Valla. Financial
support of the Ministry of Science, Education and Sports of the Republic
of Croatia through projects No. 035-0352828-2840 and No. 098-0352828-2836,
as well as of DFG through the project {}``Two dimensional cluster
lattices on graphene moirés'' is acknowledged. J. C. thanks the Alexander
von Humboldt foundation for a research fellowship. 
\end{acknowledgments}

\begin{thebibliography}{34}
\bibitem[{hyb()}]{hybrid}
\bibinfo{note}{Although the shape of the primary Dirac cone is fitted perfectly
  by the tight-binding calculations discussed later in the text, presently its
  weak hybridization with the surface state $S_1$ can not be excluded}.
\bibitem[{dop()}]{doping}
\bibinfo{note}{No conclusion about the existence of the gap at the Dirac point
  could be drawn from our sodium doping experiments, as the doping shifted the
  Dirac point below the Fermi level but altered the superstructure of graphene
  as well.}
\end{thebibliography}
 
\end{document}